# On Overcoming the Impact of Doppler Spectrum in Millimeter-Wave V2I Communications

Javier Lorca[1], Mythri Hunukumbure[2], Yue Wang[2]

[1]Telefonica I+D GCTO, Spain. e-mail: franciscojavier.lorcahernando@telefonica.com

[2]Samsung Electronics R&D Institute UK, Staines. e-mail: {mythri.h, yue2.wang}@samsung.com

*Abstract*—with the advent of 5G cellular systems there is an increased interest in exploring higher frequency bands above 6 GHz. At these frequencies, beamforming appears as a straightforward solution to overcome higher path loss thereby altering the Doppler characteristics of the received waves. Higher frequencies can suffer from strong Doppler impairments because of the linear dependency of Doppler shift with carrier frequency, which makes them challenging to use in high-mobility scenarios, particularly Vehicular-to-Infrastructure (V2I) communications. Therefore, the impact of beamforming on the Doppler characteristics of the received signals is of utter importance for future V2I systems. This paper presents a theoretical analysis of the Doppler power spectrum in the presence of beamforming at the transmit and/or the receive sides. Further approximations are made for the resulting Doppler spread and Doppler shift when the receive beam width is sufficiently small, and a possible design solution is presented to control the Doppler spread in V2I systems. The results can be of key importance in waveform and air interface design for V2I systems.

## I. INTRODUCTION

With the exponential growth in data demand for 5G cellular systems, there is an increased interest in exploring higher operational frequencies. There is a particular focus on the 6-100 GHz millimeter waves[1] frequency range (hereafter termed mm-wave), due to the potential availability of much larger bandwidths. It is well known that beamforming solutions (at both transmitter and receiver) are needed to overcome the potential path loss effects at these high frequencies, giving rise to highly directional communications.

Future V2I communications, including infotainment applications, are expected to consume very high data rates. This aspect makes mm-wave very appealing in V2I. The analysis of the behavior of Doppler shift and Doppler spread in the presence of directional beamforming is of great importance in this context. For example, such analytical results can greatly facilitate robust waveforms and air interface design against such impairments. In addition, the derived expressions can be exploited in link-level evaluations in reproducing the dynamics of realistic mm-wave channel models with beamforming. Although many publications have indicated the predicted behavior of Doppler effects in mm-wave systems [1-4], to the best of our knowledge, no analytical results or measurements on Doppler spread and Doppler shift have been published in the presence of beamforming, which are significantly different from the well-known Jakes' model. The current paper aims to fill this gap at theoretical level and apply this analysis to a practical V2I use case. We present analytical expressions and extend them to two practical cases where the receive beam width is comparable to, or much larger than, the transmit beam width. With this theoretical insight, we proceed to develop some system design guidelines for a typical V2I eMBB (enhanced Mobile Broadband) use case in high-speed trains.

The rest of the paper is organized as follows. Section II presents analytical derivations of the Doppler spectrum with beamforming, for the cases where beam width at the receiver is comparable to, or much larger than, the transmit beam width. Section III provides approximate expressions in the relevant case of having sufficiently small beam width at the receiver. In section IV, we look at a typical V2I use case in high-speed trains and provide some design guidelines to achieve a constant Doppler spread. Finally, section V is devoted to the summary and conclusions from this work.

## II. DERIVATION OF THE DOPPLER POWER SPECTRUM IN PRESENCE OF BEAMFORMING

It is well known that the Doppler shift $f_D$ of an incoming wave depends on its carrier frequency $f_c$, receiver's velocity $v$, direction of arrival $\theta$, and direction of the velocity vector $\theta_v$, according to the expression $f_D = (v/c)f_c \cos(\theta - \theta_v)$. The Doppler spectrum, expressing the power spectral density of the received waves as a function of the Doppler shift $f_d$, comprises in this case an ideally pure spectral line at $f_D$, i.e. $S(f_d) = \delta(f_d - f_D)$. If multiple waves impinge on the receiver with different angles of arrival, the Doppler shift translates into a certain

---

[1] Millimeter waves are traditionally regarded as those extending in the range of 30 GHz to 300 GHz. However, for notational simplicity the lower boundary is sometimes pushed down to 6 GHz to cope with those frequencies immediately above the ones traditionally employed in cellular communications.

Doppler spread where the Doppler spectrum contains significant contributions. When the antenna at the receiver side is omni-directional in azimuth, and in a typical urban environment where the received signal comprises the superposition of multiple waves at random directions, the well-known U-shaped Jakes Doppler spectrum is obtained [5],

$$S(f_d) = \begin{cases} \dfrac{1}{\pi f_{D\max}\sqrt{1-(f_d/f_{D\max})^2}}, & |f_d \le f_{D\max}| \\ 0, & |f_d > f_{D\max}| \end{cases} \quad (1)$$

with a maximum Doppler shift given by $f_{D\max} = (v/c)f_c$.

At higher frequencies, and particularly in the range of millimeter waves, the link budget becomes challenging because of phenomena such as signal blockage, lower antenna aperture, shadowing, or even atmospheric absorption. Beamforming arises as a straightforward way to overcome path loss by making use of transmit and receive antenna arrays, partly thanks to the smaller wavelengths that enable easier integration of multiple antennas in a tiny space. The dynamic nature of cellular links makes it crucial to employ dynamic beamforming, whereby transmit and receive beams are ideally collinear to maximize beamforming gains. In these conditions, the number of multipath components and the angular spread of the incoming waves are expected to decrease because of beamforming, which in turn reduces the Doppler spread [2].

In the following we derive expressions for the Doppler power spectrum suitable for mm-wave systems in scenarios where both the transmitter and receiver have beamforming capabilities, with half-power beam widths in the horizontal plane given by $\theta_H^{TX}$ and $\theta_H^{RX}$, respectively. In what follows we assume that the transmitter is the mm-wave base station and the receiver is the user device, but the same conclusions would apply by reversing the roles of the two ends.

In the presence of beamforming at both sides of the link only a fraction of the multipath components is captured by the receiver, namely those with angles of arrival between $-\theta_H^{RX}/2$ and $+\theta_H^{RX}/2$ in the horizontal plane (taking the origin of angles at an imaginary line connecting the transmitter and the receiver). We assume a relative velocity vector $v$ in the horizontal plane characterized by an angle $\theta_v$ (Fig. 1).

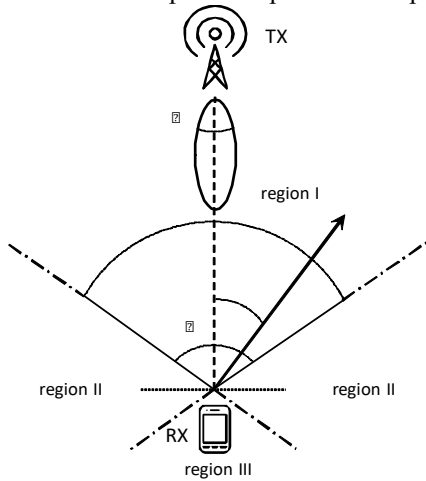

Fig. 1. Geometric definition of the magnitudes $\theta_H^{TX}$ (horizontal beam width at the TX side), $\theta_H^{RX}$ (horizontal beam width at the RX side) and $\theta_v$ (angle between the velocity vector and the line of sight between TX and RX).

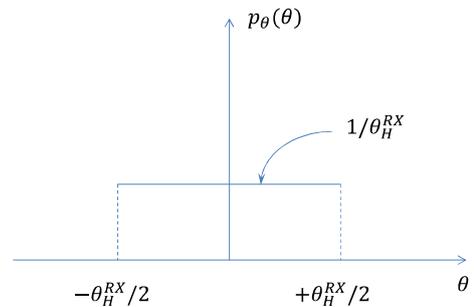

Fig. 2. Probability density function of the angle of arrival $\theta$ when the receive beam width is comparable to the transmit beam width.

Assuming the presence of random scatterers within the transmit beam width, and that these scatterers can alter the angles of the multipath components in an unpredictable way, we can assume that the angles of arrival corresponding to the waves coming from a given scatterer follow a uniform distribution in the range $[-\theta_H^{RX}/2, +\theta_H^{RX}/2]$, as illustrated in Fig. 2 and stated by the following equation:

$$p_\theta(\theta) = \dfrac{1}{\theta_H^{RX}} \cdot rect\left(\dfrac{\theta}{\theta_H^{RX}}\right). \quad (2)$$

It is reasonable to consider such a uniform probability of arrival angles if the receive beam width is not very large compared to the transmit beam width. If this condition holds, most of the reflected paths will be caused by a few scatterers under the visibility of the receiver's antennas, and reflections will be likely spread following a simple uniform distribution. If, however, the receive beam width is much larger than the transmit beam width, the chances increase that reflections are grouped in the form of angular clusters separated by relatively empty spatial regions [7]. The narrow beams present at the transmit side prevent the signals from being scattered into all directions in space, following instead more structured spatial patterns that could be captured by the receiver if $\theta_H^{RX}$ is wide enough.

Considering a uniform distribution of arrival angles inside each cluster is probably not an entirely solid assumption [6-7]. However, recent measurements confirm such a uniform distribution of arrival angles at 60 and 70 GHz [8]. Moreover, it simplifies the analysis and can be used as a good starting point for further studies. In what follows we analyze both cases by utilizing a common theoretical approach as detailed in sections II.A and II.B, which will lead us to several important conclusions in Section III.

*A. Case when the receive beam width is comparable to the transmit beam width*

Upon reception of a multipath component with angle $\theta$,

the instantaneous frequency will be shifted by an amount given by $f_d = (v/c) f_c \cos(\theta - \theta_v)$. Hence, the probability density function (pdf) of the Doppler shift can be calculated by:

$$p_{f_d}(f_d) = p_\theta(\theta)\left|\frac{d\theta}{df_d}\right| = p_\theta\left\{\theta_v + \cos^{-1}\left(\frac{c}{vf_c}f_d\right)\right\}\left|\frac{d\theta}{df_d}\right|$$

$$= \frac{rect\left\{\frac{1}{\theta_H^{RX}}\left(\theta_v + \cos^{-1}\left(\frac{f_d}{f_{D\max}}\right)\right)\right\}}{\theta_H^{RX} f_{D\max}\sqrt{1 - \left(\frac{f_d}{f_{D\max}}\right)^2}}. \quad (3)$$

The resulting expression is hence non-zero only for the range $[f_{d,1}, f_{d,2}]$, where three spatial regions can be defined depending on the relative values of $\theta_v$ and $\theta_H^{RX}$ (see Figure 1):

- *Region I*, $|\theta_v| \leq \frac{\theta_H^{RX}}{2}$: the range in which the pdf of the Doppler shift is non-zero is given by:

$$f_{d,1} = f_{D\max} \cos\left(\frac{\theta_H^{RX}}{2} + |\theta_v|\right),$$
$$f_{d,2} = f_{D\max}. \quad (4)$$

The maximum Doppler shift ($f_{D\max}$) is achieved in this case when the received multipath component is collinear with the velocity vector.

- *Region II*, $\frac{\theta_H^{RX}}{2} < |\theta_v| \leq \pi - \frac{\theta_H^{RX}}{2}$: the corresponding boundaries are given by:

$$f_{d,1} = f_{D\max} \cos\left(\frac{\theta_H^{RX}}{2} + |\theta_v|\right),$$
$$f_{d,2} = f_{D\max} \cos\left(\frac{\theta_H^{RX}}{2} - |\theta_v|\right). \quad (5)$$

- *Region III*, $\pi - \frac{\theta_H^{RX}}{2} < |\theta_v| \leq \pi$: the boundaries are given by:

$$f_{d,1} = -f_{D\max},$$
$$f_{d,2} = f_{D\max} \cos\left(\frac{\theta_H^{RX}}{2} - |\theta_v|\right), \quad (6)$$

with the minimum Doppler shift ($-f_{D\max}$) being achieved when the received multipath component is collinear with the velocity vector.

With these expressions, it is possible to define the magnitudes:

$$f_{d,shift} = \frac{f_{d,1} + f_{d,2}}{2},$$
$$\Delta f_d = f_{d,2} - f_{d,1}, \quad (7)$$

which represent the central Doppler shift and the Doppler spread of the resulting Doppler power spectrum, respectively. The pdf in (3) can then be re-written as:

$$p_{f_d}(f_d) = \frac{rect\left\{\frac{1}{\Delta f_d}(f_d - f_{d,shift})\right\}}{\theta_H^{RX} f_{D\max}\sqrt{1 - \left(\frac{f_d}{f_{D\max}}\right)^2}}, \quad (8)$$

where $f_{d,shift}$ and $\Delta f_d$ can be calculated by means of (4)-(6) for the three spatial regions.

The Doppler power spectrum expresses the power spectral density as a function of the Doppler shift $f_d$ at the receiver. Assuming that the scattering environment is predominantly reflective (with very little diffuse reflection), and that the transmitter's antenna pattern is sufficiently flat within its horizontal beam width $\theta_H^{TX}$, the Doppler power spectrum can be considered proportional to the pdf of the Doppler shift further affected by the receiver's antenna gain $G(\theta)$ (expressed as a function of the Doppler shift). Hence, we can write:

$$S(f_d) = p_{f_d}(f_d) G(\theta)$$
$$= \frac{G\left\{\theta_v + \cos^{-1}\left(\frac{f_d}{f_{D\max}}\right)\right\}}{\theta_H^{RX} f_{D\max}\sqrt{1 - \left(\frac{f_d}{f_{D\max}}\right)^2}} \cdot rect\left\{\frac{1}{\Delta f_d}(f_d - f_{d,shift})\right\}. \quad (9)$$

Note that the above expression does not depend on $\theta_H^{TX}$, but this only comes from the assumption that arrival angles follow a uniform probability distribution as in (2). In other environments, this condition would not hold and the analysis would lead to much more complicated expressions. However, the above assumption can be representative of a large number of practical scenarios where transmitter and receiver are not too close to each other, and the receive beam width is not very large compared to the transmit beam width.

If the receive antenna gain is sufficiently flat within the receive beam width, the Doppler spectrum will comprise a "portion" of the classical U-shaped Jakes spectrum contained between the Doppler frequencies $f_{d,1}$ and $f_{d,2}$ (Fig. 3). Eventually one of the ends of the Doppler spectrum can become $+f_{D\max}$ or $-f_{D\max}$ when there is a chance that one multipath component is collinear with the velocity vector.

### B. Case when the receive beam width is much larger than the transmit beam width

In this case, there may likely appear multiple clusters of incoming waves at the receiver caused by surrounding scatterers. The lower predominance of diffuse reflection compared to lower frequencies, together with the small beam width of the transmissions, increase the chances that receiving waves appear in the form of clusters presenting approximately uniform probability distributions over certain angles. In these conditions, we may generalize the Doppler power spectrum in (9) to comprise a sum of terms $S(f_d) = \sum_{j=0}^{M-1} S_j(f_d)$, where $M$ is the number of receiving clusters and $S_j(f_d)$ takes the form:

$$S_j(f_d) = \frac{G\left\{\theta_v + \cos^{-1}\left(\frac{f_d}{f_{D\max}}\right)\right\}}{\theta_H^{RX} f_{D\max} \sqrt{1 - \left(\frac{f_d}{f_{D\max}}\right)^2}} \cdot rect\left\{\frac{1}{\Delta f_d^j}\left(f_d - f_{d,shift}^j\right)\right\}. \quad (10)$$

In (10) the Doppler shift and Doppler spread values of the $j$-th cluster of incoming waves are denoted as $f_{d,shift}^j$ and $\Delta f_d^j$, respectively. Fig. 4 very schematically illustrates the resulting Doppler power spectrum in a hypothetical case with $M = 3$ (typical numbers for $M$ are between 1 and 5, as reported in [7]). It is apparent here that the overall Doppler width largely depends on the number of clusters and the relative orientations of the incoming waves with respect to the velocity vector.

TABLE I: SUMMARY OF APPROXIMATE DOPPLER SHIFT AND DOPPLER SPREAD VALUES WHEN THE RECEIVE BEAM WIDTH IS SUFFICIENTLY SMALL

| Angular region | Doppler shift | Doppler spread |
|---|---|---|
| $\lvert\theta_v\rvert \leq \dfrac{\theta_H^{RX}}{2}$ | $f_{D\max}\left(1 - \dfrac{\theta_H^{RX}\lvert\theta_v\rvert}{4}\right)$ | $f_{D\max}\dfrac{\theta_H^{RX}}{2}\lvert\theta_v\rvert$ |
| $\dfrac{\theta_H^{RX}}{2} < \lvert\theta_v\rvert \leq \pi - \dfrac{\theta_H^{RX}}{2}$ | $f_{D\max}\cos\theta_v$ | $f_{D\max}\theta_H^{RX}\lvert\sin\theta_v\rvert$ |
| $\pi - \dfrac{\theta_H^{RX}}{2} < \lvert\theta_v\rvert \leq \pi$ | $-f_{D\max}\left(1 - \dfrac{\theta_H^{RX}\lvert\pi - \theta_v\rvert}{4}\right)$ | $f_{D\max}\dfrac{\theta_H^{RX}}{2}\lvert\pi - \theta_v\rvert$ |

### III. APPROXIMATE EXPRESSIONS WHEN THE RECEIVE BEAM WIDTH IS SUFFICIENTLY SMALL

From (5) – (7) it is apparent that, if the receive beam width $\theta_H^{RX}$ is sufficiently small, an easier treatment can be given to the expressions in section II.A by considering the approximations $\cos(\theta_H^{RX}/2) \cong 1$ and $\sin(\theta_H^{RX}/2) \cong \theta_H^{RX}/2$.

We can also state that $\cos(\lvert\theta_v\rvert) \cong 1$, $\sin(\lvert\theta_v\rvert) \cong \lvert\theta_v\rvert$ in Region I, and $\cos(\lvert\theta_v\rvert) \cong -1$, $\sin(\lvert\theta_v\rvert) \cong \lvert\pi - \theta_v\rvert$ in Region III. These approximations allow easy derivation of the approximate Doppler shift and Doppler spread values. Table I summarizes such results for the three above defined spatial regions. Doppler spread is significantly reduced compared to that in classical Jakes spectrum, and Doppler spectrum resembles an ideal Doppler shift quite closely because of beamforming. This makes it much easier for the receiver to track and compensate the effects of mobility. The expressions in Table I remain valid even if the distribution of arrival angles is not uniform, because Doppler shift and Doppler spread values only depend on the angular region where the Doppler power spectrum is non-zero, and this in turn is determined by $\theta_H^{RX}$.

Technically, Doppler shift can be quite easily accommodated by re-alignment if the amount of shift can be estimated at the receiver. Doppler spread is a much more serious problem, particularly for multi-carrier systems, as the subcarrier spacing has to be designed to cater for the maximum Doppler spread. From the previous results it is clear that Doppler spread at a given constant velocity is $O(f_c \cdot \theta_H^{RX})$, whilst at a given constant frequency is $O(v \cdot \theta_H^{RX})$. This is particularly interesting as $\theta_H^{RX}$ can be made inversely proportional to the carrier frequency or the receiver velocity, by increasing the number of antennas in the horizontal dimension. In a system spanning a wide range of mm-wave frequencies, the number of antennas can be increased with frequency to overcome path loss as well as maintain a constant Doppler spread. In a system with narrow frequency span but dealing with a wide range of receiver velocities, the parameter $\theta_H^{RX}$ can be used again to control the Doppler spread. This can be extremely useful in the design of V2I mm-wave systems as we explain in the use case below.

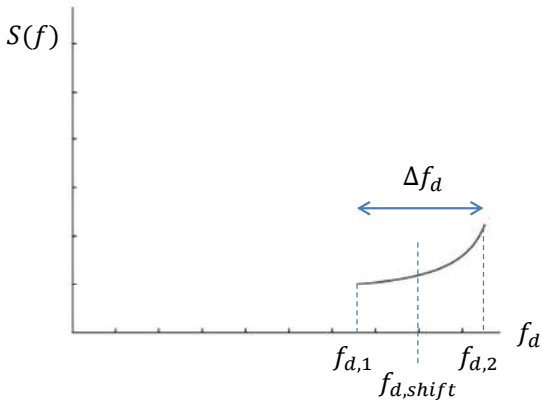

Fig. 3. Schematic illustration of the Doppler power spectrum when the receive beam width is comparable to the transmit beam width.

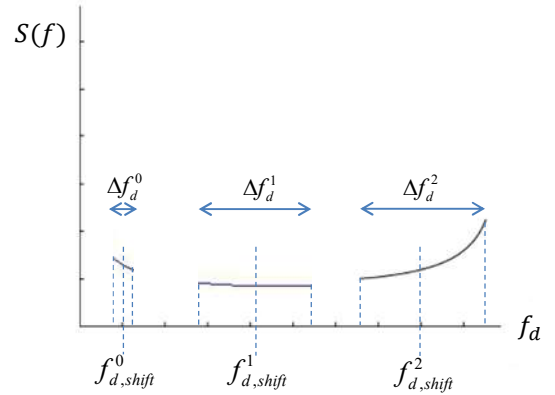

Fig. 4. Schematic illustration of the Doppler power spectrum when the receive beam width is much larger than the transmit beam width; example with $M = 3$ clusters.

## IV. DESIGN PARAMETERS FOR HIGH SPEED TRAINS

High-speed trains are a classic example where extreme data rates and capacity will need to be delivered under challenging Doppler conditions in future 5G systems. Although there have been continuous improvements in wireless quality of service (QoS) provision for railways, from GSM-R systems to currently developing LTE-R [9], future traffic demands from this group mobility scenario will make 5G mm-wave systems very appealing for high speed trains. Although beam tracking and alignment may seem very challenging at these extremely high speeds (up to 500 km/h), the velocity of the train at each point of the track can be predicted to a very high accuracy. There is advanced research currently on-going in this area [10] and we believe that the technologies will evolve to enable beam tracking and alignment at high speed. When the high speed trains are served by base stations arranged along the side of the track, beam alignment is most challenging when the train is moving nearly perpendicular to the base stations. An associated problem will be accommodating for the very large range of Doppler spreads, when the train speed varies from stationary conditions up to 500 km/h as it travels between stations.

V2I communications covers both uplink and downlink directions. Focusing for simplicity only on the downlink (from the infrastructure to the train), which will have significantly higher data rate requirements, we propose to control the receiver's (mounted on the train) horizontal beam width $\theta_H^{RX}$. By making $\theta_H^{RX}$ inversely proportional to the train speed it is apparent that near-constant Doppler spread can be achieved. The antenna array mounted on the train can have multiple elements (or panels), which can be electronically combined/detached to yield the desired beam width. When the train is departing from the train station the value of $\theta_H^{RX}$ will be at its maximum (but still small, as discussed in section III). The receiver antenna gain will be smaller in this area, and the base stations will have to be located at shorter distances to compensate for the more challenging link budget. However, given that the train velocity is still small, there would not be a significant reduction in the time between handovers from one base point to another. As the train speed picks up, the receiver beam width can be progressively reduced to maintain a constant maximum Doppler spread. This also means that the receiver antenna gains are increased, and the base stations are spaced further apart, allowing sufficient time between handovers. Having a Doppler spread that is constant (or only changes within a manageable range) allows the design of multi-carrier systems (like OFDM) that do not have to over-compensate for a very large range of Doppler spreads. In essence, this enables the spectral efficiency of the system to be increased, and the transceivers to be much simpler, dealing with only a single value of the subcarrier spacing.

As a very simple example of the magnitudes involved, let us assume a 28 GHz link between a moving train and the fixed infrastructure that is offering mm-wave connectivity to the train. The speed limit could be set to 500 km/h, hence yielding a maximum Doppler shift of ±12.96 kHz, which can be easily predicted and compensated along the train's trajectory. The Doppler spread however poses more challenges to the receiver and should be kept under control. From Table I it is apparent that the worst case value of the Doppler spread is $f_{D\max}\theta_H^{RX}$. Assuming a 10º beam width at low speeds (e.g. those below 50 km/h), the beam width can be progressively decreased to 1º as the train reaches its peak speed of 500 km/h. This can be achieved by changing the receiver beam width according to the following expression, where the units have been changed for convenience:

$$\theta_H^{RX}(º) = \frac{1.4 \times 10^4}{f_c(GHz) \cdot v(km/h)}. \quad (11)$$

The different values of $\theta_H^{RX}$ can be realized by combining more or less numbers of antennas elements at the train-mounted array according to the train velocity along the different parts of the trajectory. This simple strategy can keep the maximum Doppler spread under control and, thus, allow the use of a single subcarrier spacing for the wide range of velocities that are a characteristic of high-speed trains.

The diagram in Fig. 5 schematically illustrates this scenario for the two speed extremes. The diagram is not drawn to scale.

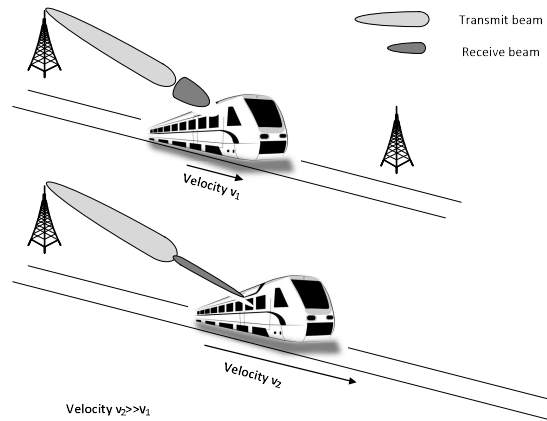

Fig. 5. Schematic diagram showing the receiver beam width adjustment to control the maximum Doppler spread.

## V. Summary and conclusions

This paper presents a theoretical analysis on the Doppler spread and Doppler shift in the presence of directional transmission and reception in the mm-wave spectrum, and applies it to a V2I application in 5G. The analysis is built on the assumption of clustered reception of incoming waves as observed in recent measurement campaigns. Further approximations resulting in simple expressions of Doppler spread and Doppler shift are also presented (Table I). The work analytically shows that the Doppler spread is much confined to a single narrow segment when the receiver beam width is sufficiently small, and to multiple confined segments when the receiver beam width is much larger than the transmitter beam width. These results can encourage researchers to conduct further experiments on Doppler spectrum, which can be of key importance both for air-interface evaluations and in waveform design at higher frequencies.

One of the key observations from this analysis is that the Doppler spread can be effectively controlled by adapting the receiver beam width. This enables system designers to maintain a fixed and manageable subcarrier spacing in OFDM-like multi-carrier systems. We illustrate this by means of a mm-wave V2I example with high speed trains. We hope that this analysis will encourage further work in this area, thus yielding more insights for mm-wave system design in high speed vehicular applications.

## Acknowledgment

The research leading to these results received funding from the European Commission H2020 programme under grant agreement nº 671650 (5G PPP mmMAGIC project).